# Chromium isotope evidence in ejecta deposits for the nature of Paleoproterozoic impactors

**Bérengère Mougel[1], Frédéric Moynier[1, 2], Christa Göpel[1], and Christian Koeberl[3]**

[1]*Institut de Physique du Globe de Paris, Université Sorbonne Paris Cité, CNRS UMR 7154, Paris, France*

[2] *Institut Universitaire de France and Université Paris Diderot, Paris, France*

[3] *Department of Lithospheric Research, University of Vienna, Althanstrasse 14, A-1090 Vienna, Austria and Natural History Museum, Burgring 7, A-1010 Vienna, Austria.*

Corresponding: Mougel@ipgp.fr

**ABSTRACT**

Non-mass dependent chromium isotopic signatures have been successfully used to determine the presence and identification of extra-terrestrial materials in terrestrial impact rocks. Paleoproterozoic spherule layers from Greenland (Grænsesø) and Russia (Zaonega), as well as some distal ejecta deposits (Lake Superior region) from the Sudbury impact (1,849±0.3 Ma) event, have been analyzed for their Cr isotope compositions. Our results suggest that 1) these distal ejecta deposits are all of impact origin, 2) the Grænsesø and Zaonega spherule layers contain a distinct carbonaceous chondrite component, and are possibly related to the same impact event, which could be Vredefort (2,023±4 Ma) or another not yet identified large impact event from that of similar age, and 3) the Sudbury ejecta record a complex meteoritic signature, which is different from the Grænsesø and Zaonega spherule layers, and could indicate the impact of a heterogeneous chondritic body.

# 1. INTRODUCTION

The Earth has been subjected to numerous large impacts since its accretion (i.e., the giant impact related to the origin of the Moon, the Late Heavy Bombardment), but little evidence of its bombardment history is preserved in modern geologic records (e.g., Koeberl, 2006a,b). About 190 impact structures have so far been confirmed on the Earth's surface, but very few date to the Paleoproterozoic or older. In contrast to remote sensing and other geophysical methods, the confirmation of impact structures on Earth requires the detection of either shock metamorphic effects in minerals and rocks, and/or the presence of a meteoritic component in these rocks. Apart from studying meteorite impact craters directly, information can also be gained from the study of impact ejecta. These are layers of melted and shocked rock or mineral fragments, including millimeter- to centimeter-sized impact spherules and glasses (such as tektites) that form from melt and vapor condensate droplets, as well as accretionary lapilli (Glass and Simonson, 2012). In the absence of meteorite fragments, the presence of a meteoritic component within the target rocks can be verified by measuring abundances and inter-element ratios of siderophile elements (e.g., Cr, Co, Ni), and especially the Platinum Group Elements (PGE), which are orders of magnitudes more abundant in meteorites than in terrestrial crustal rocks (Koeberl et al., 2012). The Re-Os isotopic method is also traditionally used for the detection of iron meteorite and chondritic material because they have a different $^{187}Os/^{188}Os$ ratio from the Earth's crust (Koeberl, 2014). However, all these methods are not sufficient to distinguish between chondrite types. The Cr isotope method allows a better identification of the type of meteoritic material involved because well-resolved Cr isotopic differences do exist between meteorites (Shukolyukov and Lugmair, 1998; Trinquier et al., 2007, Moynier et al., 2009). The Cr isotopic composition of each chondrite group is distinct; and while the $^{54}Cr/^{52}Cr$ ratio of some achondrites overlaps with that of chondrites (e.g., eucrites and ordinary chondrites, (Fig. 1) (Trinquier et al., 2007), it is possible to distinguish them from one another by coupling $^{53}Cr/^{52}Cr$ and $^{54}Cr/^{52}Cr$ ratios (Fig. 2). This approach has been successfully used for the identification of the impactors involved in the formation of the Morokweng, Bosumtwi, Clearwater, Lappajärvi, and Rochechouart (Koeberl et al., 2007) impact structures, as well as for ancient ejecta layers (e.g., Trinquier et al., 2006; Quitté et al., 2007; Kyte et al., 2003, 2011).

Evidence for the bombardment of the Earth (i.e., impact structures and ejecta) between 1.6 Ga and 2.5 Ga is rare. Only three distal impact ejecta layers, namely the Grænsesø (South Greenland) and Zaonega (Karelia, North West Russia) spherule layers, and the Sudbury layer in the lake Superior region ejecta (North America) have been recognized. The ages of these deposits range from 1,830±3 Ma to 2,130±65 Ma, and brackets the ages of the two largest and oldest terrestrial impact structures presently found at the surface of the Earth, Vredefort (2,023±4 Ma; Kamo et al., 1996) and Sudbury (1,849±0.3 Ma; Krogh et al., 1984) (Turtle et al., 2005). However, there is no geochemical evidence that links the ejecta in Greenland and Russia to one of these events, or other large undiscovered or totally eroded impact event(s). Here, we investigate the Cr isotopic composition of samples from these ejecta layers in order to identify the nature of the impactors, as well as to discuss the possible relationships between these layers, and their link with the Sudbury and Vredefort impact structures. Specifically, we analyzed orphan spherule layers from Greenland and Russia, and confirmed distal ejecta deposits (Lake

Superior region, Canada) from the Sudbury impact event (Chadwick et al., 2001; Koeberl et al., 2002; Huber et al., 2014a; Huber et al., 2014b; Petrus et al., 2015). Our results provide new evidence for an impact origin of these ejecta deposits, and therefore, the sources of bombardment of the Earth at around 2 Ga.

## 2. SAMPLES AND METHODS

### 2.1 Paleoproterozoic ejecta layers

Layers interpreted as ejecta from the Sudbury event have been recognized at more than 15 sites in the Lake Superior area (e.g., Addison et al., 2005; Cannon et al., 2010). Many arguments support the idea that they are all products of a single impact: 1) Their similarities in geological characteristics; 2) the regional persistence of the layers at a constant stratigraphic level atop the main local iron formations (Fig. S1); 3) their major and trace element chemical composition closer to that of the "Onaping" melts in Sudbury structure than to any local rocks in the Lake Superior region (Cannon et al., 2010); 4) the regional variations in thickness and petrographic content of the layers consistent with their distances from the current crater location (Cannon et al., 2010); 5) the well-known age of the Sudbury impact close to the estimated depositional age of the layers; and 6) the fact that no other contemporaneous impact structure has yet been found, neither other ejecta layer at any of the sites where the Sudbury ejecta layer was already observed.

The detection of a meteoritic component in rocks from Sudbury is supported by the geochemical studies of both the crater's rocks and ejecta layers (e.g., Pufahl et al., 2007; Cannon et al., 2010). More geochemical evidence for the composition of the Sudbury impactor recently came from the signature of PGE advocating a meteoritic, and more specifically a chondritic origin (Huber et al., 2014b; Petrus et al., 2015).

The Grænsesø spherule layer is the least studied layer among the three described here. It is located in the Ketilidian orogeny (South Greenland) and composed of spherules within a dolomite layer that constitutes the upper part of the Paleoproterozoic metasedimentary Vallen group. Spherules represent more than 15% of the total volume of the layer, at least locally the rest being carbonates, chert clasts, and epiclastic sand grains (Chadwick et al., 2001). Despite the absence of evidence for shock features, Chadwick et al. (2001) re-interpreted their origin based on a detailed textural analysis of individual spherules, and presented evidence for an impact origin rather than resulting from a volcanic or biological activity. The spherules are generally larger (~1 mm) than possible spherulitic fossils (~0.3 mm), and their shapes are more circular than volcanic spherules that tend to be on average more elongated (Heiken and Vohletz, 1985). Provided that the impact origin is confirmed, this layer must have been associated with a large impact event because of the high abundance of spherules and estimated thickness of the layer (Chadwick et al., 2001). Based on the ages of the intrusions that crosscut the basement and the ejecta layer, the age of the layer is loosely constrained between 1,848±3 Ma and 2,130±65 Ma (Chadwick et al., 2001; Garde et al., 2002). This time interval is concordant with the ages of both the Vredefort and Sudbury impact structures but too broad to infer a direct link with one of them. Moreover, there is no geochemical evidence yet for the presence of a meteoritic component in this layer. The only bulk rock composition published so far (Chadwick et al., 2001) shows very slight PGE enrichments compared to the average composition of the continental crust, and PGE patterns different from those of Sudbury ejecta (Fig. 3).

Recently, a similar spherule layer was discovered in the Paleoproterozoic Zaonega formation in Karelia (North West Russia), which represents supplementary physical evidence for a ~2 Ga impact event. The spherules are enriched and distributed into multiple layers in the drill cores, suggesting that the ejecta were disturbed during and/or after its initial deposition (Huber et al., 2014a). This spherule layer shows structural similarities with the Grænsesø spherule deposits. Like the Grænsesø layer, this layer is hosted in dolostones, whose depositional age is constrained between 1.98±0.02 Ga and 2.05±0.02 Ga (Puchtel et al., 1998; Hannah et al., 2008). The spherules are circular with an average diameter of 0.8 mm, and are surrounded by a cement assemblage similar to the one found in the Grænsesø layer. Huber et al. (2014a) have shown that, although the PGE ratios and concentrations of these ejecta do not allow the precise identification of the nature of the projectile, they are clearly distinct from magmatic samples and contain a chondritic component. Platinum Group Element abundances and ratios of this layer are different from the Sudbury ejecta compositions, but on the contrary share similarities with the Vredefort granophyre composition (Fig. 3). However, while earlier studies based on Re-Os isotopes analysis of impact melts rocks from the Vredefort structure also confirmed the presence of a meteoritic component (≤0.2%), Cr isotope investigation was not sufficiently sensitive to pinpoint the meteoritic component, and therefore, to identify the impactor's nature (Koeberl et al., 1996, 2002). The discovery of the possible ejecta layers in Greenland (Chadwick et al., 2001) and Russia (Huber et al., 2014a), if confirmed, thus open new perspectives for the detection and identification of extra-terrestrial material during this geological period, and for evaluation of their origin and relationship with the Sudbury and Vredefort impact events.

In this study, we measured the Cr isotope compositions of four bulk ejecta samples from the Sudbury ejecta layer (C9, C18, PR9, and PR10), two from the Grænsesø spherule layer (GL-1 and GL-5), and two from the Zaonega spherule layer (FD27.36 and FD26.51), as well as a terrestrial basalt standard, BHVO-2. The corresponding petro-geochemical descriptions, locations, as well as major, trace, and PGE data are given in Huber et al. (2014b), Chadwick et al. (2001), and  Huber et al. (2014a), respectively.

## 2.2. Analytical procedure

Thirty mg of bulk rock samples were entirely dissolved in a mixture of concentrated HF and $HNO_3$ in Teflon bombs at 140°C for several days, including multiple ultrasonication steps. No residual gels and/or refractory phases were observed during the final inspection of the solution. This dissolution procedure has been tested with success on pure chromite grains. The chemical procedure adopted from Trinquier et al. (2008) includes 3 separation steps of Cr on cationic exchange resin AG50W-X8, and allows a ≥99% yield (Birck and Allègre, 1984) (Table S1). All chemical separations and isotopic measurements were performed at the Institut de Physique du Globe de Paris, France. $^{53}Cr/^{52}Cr$ and $^{54}Cr/^{52}Cr$ isotope ratios were measured by multi-collection (9 cups) Thermal-Ionization Mass-Spectrometry (TIMS) Fisher Scientific Triton, Cr concentrations by ICP-MS-HR Fisher Scientific Element 2. Details of the Cr isotopic measurements were published by Göpel et al. (2015).  Purified Cr samples were loaded in chloride form on degassed W filaments together with an Al-silicagel-$H_3BO_3$ emitter in order to facilitate Cr emission and stability. A typical measurement comprises 20 blocks of 20 cycles. Each single run is a combination of 3 successive multi-collection measurements (Table S2) in static mode with $^{53}Cr$, $^{52}Cr$ and $^{54}Cr$ isotopes shifted by one mass unit in the center cup and optimized using the zoom optics to adjust the peaks shape and centering. The isotopic ratios obtained with each configuration represent independent measurements that can be compared to each other, and are used to control the evolution of the instrument over time. In parallel, we dissolved and measured a terrestrial rock standard (BHVO-2) to ensure data accuracy. $^{53}Cr/^{52}Cr$ and $^{54}Cr/^{52}Cr$ ratios of each beam configuration were normalized using an exponential law to $^{52}Cr/^{50}Cr = 19.28323$ (Shields, 1966). The data are not second-order normalized using $^{54}Cr/^{52}Cr$ as it has been the case in some previous papers (e.g., Lugmair and Shukolyukov 1998; Koeberl et al., 2007). $^{56}Fe$ was monitored in order to control the possible isobaric interference of $^{54}Fe$ with $^{54}Cr$. When possible, samples were loaded at least two times on different filaments. The total number of measurements is given by the letter $n$ in the Table 1. No full procedure duplicate could be made because of the limited amount of samples available. Final Cr isotopic data are given in ε-units that represent the relative deviation in parts per 10,000 of $^{53}Cr/^{52}Cr$ and $^{54}Cr/^{52}Cr$ ratios from a terrestrial standard (NIST SRM 3112a Cr standard). At a scale of 1-2 weeks measurement sessions, the external reproducibility (2 sd) was measured and turn out to be on average, 9 ppm and 20 ppm for $^{53}Cr/^{52}Cr$ and $^{54}Cr/^{52}Cr$, respectively.

## 3. RESULTS

The results are presented in Table 1, together with complementary data from the literature. The basalt BHVO-2 shows typical Cr terrestrial values. Sudbury ejecta (PR9, PR10, C9 and C18) exhibit large variations in Cr isotopic compositions ranging from -0.20±0.09 to 0.26±0.07 for $\varepsilon^{54}Cr$ and from -0.04±0.05 to 0.19±0.04 for $\varepsilon^{53}Cr$. A clear distinction in the geochemical composition is observed between the samples from Pine River (PR) and Coleraine (C) drill cores. The average Cr concentration of PR samples is higher than the one of C samples. This is also true for the PGE abundances (Huber et al., 2014b) measured in Pine River rocks, which show slightly larger enrichments than Coleraine rocks. The opposite is observed for Cr isotopic compositions. C18 exhibits a terrestrial Cr isotope signature ($\varepsilon^{53}Cr=0.02\pm0.03$;

$\varepsilon^{54}$Cr=0.05±0.09) and is characterized by lower Cr and other siderophile element concentrations. C9 is more enriched in Cr, and shows positive isotope anomalies ($\varepsilon^{53}$Cr=0.19±0.04; $\varepsilon^{54}$Cr=0.26±0.07). PR10 is the most enriched in PGE, Cr, Ni, and Co (Table1), but exhibits no Cr isotope anomaly ($\varepsilon^{53}$Cr=-0.03±0.04; $\varepsilon^{54}$Cr=-0.01±0.08), whereas PR9 has a small but significant $^{54}$Cr deficit (-0.20±0.09) at almost similar siderophile and PGE contents.

The Cr isotope compositions of the samples GL-1 and GL-5 from the Grænsesø spherule layer are $\varepsilon^{53}$Cr=0.14±0.05; $\varepsilon^{54}$Cr=0.54±0.11 and $\varepsilon^{53}$Cr=0.07±0.04; $\varepsilon^{54}$Cr=0.35±0.09, respectively. These values are lower than samples FD27.36 and FD26.51 from the Zaonega spherule layer ($\varepsilon^{53}$Cr=0.18±0.04; $\varepsilon^{54}$Cr=1.26±0.09 and $\varepsilon^{53}$Cr=0.14±0.06; $\varepsilon^{54}$Cr=1.06±0.11 respectively). Their siderophile element abundances are also distinct, with FD27.36 having slightly lower Cr and strongly higher Co and Ni concentrations than GL-5 (Table 1). FD samples show PGE compositions that are different from those of PR and C samples (Fig. 3). No PGE concentrations have been measured yet in the GL samples. However, the unique PGE bulk data (GGU71380, Fig. 3) for the Greenland spherule layer shows lower PGE abundances (except for Rh and Ru) than for FD samples, but similar characteristics in their HSE patterns (i.e. increasing abundances with siderophility, and Pt negative and Rh positive anomalies).

## 4. DISCUSSION

### *4.1. Sudbury ejecta deposits*

Several studies have attempted to identify the nature of the Sudbury projectile (e.g., Morgan et al., 2002; Pufahl et al., 2007; Grieve et al., 2010), but only with moderate success. Recently, Huber et al. (2014b) and Petrus et al. (2015) arrived at the same conclusion using the PGE approach applied on two different sample sets (distal ejecta deposits and impactites from the crater), which is that the object that hit the Earth at Sudbury is characterized by a chondritic composition. However, the authors were not able to differentiate between an ordinary and an enstatite chondrite because inter-PGE ratios of both types of meteorites are too similar. This Cr isotope investigation represents another independent and complementary approach that attempts to resolve the nature of the Sudbury impactor.

Our results for Lake Superior region ejecta samples (C9, C18, PR9, PR10) show a wide range of $\varepsilon^{53}$Cr and $\varepsilon^{54}$Cr values, from positive to negative signs (Table 1, Fig. 1&2). The Cr isotope composition and the low siderophile element abundances of C18 unambiguously show a terrestrial signature.

PR10 also exhibits a terrestrial Cr isotope composition, but an enriched siderophile element content that made it initially a good candidate for the detection of the impactor. This absence of Cr isotopic anomalies for PR10 despite its clear siderophile element enrichments can be interpreted into different ways. On the one hand, the enrichments in PGE can either be explained by an input from the impactor, or by preferential remobilization and re-concentration of PGE during igneous processes related to the impact (e.g., uptake by specific minerals such as sulfides). On the other hand, a terrestrial-like Cr isotope composition suggests either the absence of meteoritic Cr in this rock, or possibly the presence of an enstatite chondrite component (i.e., ≤3-4% to account for PR10 Cr signature). Huber et al. (2014b) previously showed that the PGE in the PR ejecta experienced some degree of mobilization, but that these rocks conserved Ru/Ir, Rh/Ir and Pt/Ir ratios that are consistent with a chondritic source. Therefore, unless PGE and Cr are totally decoupled, the contribution of an enstatite chondrite component in PR10 is compatible with its measured Cr composition. In addition, Petrus et al. (2015) also proposed based on the PGE ratios, the occurrence of enstatite (or possibly ordinary) chondrite material in the rocks from the crater. Nevertheless, the possibility that all Cr in PR10 is from a terrestrial source despite PGE, Ni, Co enrichments cannot be totally excluded by our results.

One way to test this hypothesis was to measure an additional sample from the same area for which chondritic PGE enrichments have been observed too. This sample (PR9) exhibits a negative $\varepsilon^{54}$Cr value (-0.20±0.09), which is distinct from both the range of terrestrial and enstatite chondrite compositions, thus in contradiction with PR10 composition. Because only chondritic material can account for the PGE compositions and trends observed in the Sudbury impactites (Huber et al., 2014b and Petrus et al., 2015), the unique candidate left over and that is characterized by negative $\varepsilon^{54}$Cr values are ordinary chondrites (Fig. 1). The $^{54}$Cr deficit of PR9 is relatively small (-0.20±0.09), however anomalies of similar size have already been observed and used for the interpretation of other impact structures such as Morokweng and Lappajarvi (e.g., Koeberl et al., 2007). The isotope anomalies in impact melt rocks are usually small because most of the chondrite types (except carbonaceous chondrites) show very small Cr isotopic variations compared to Earth (< 1$\varepsilon$), and also because they are largely diluted in terrestrial material during the impact afterwards.

First order mixing calculations can be useful to test the type of impactor that is compatible with the Cr isotopic anomalies observed in our samples. Therefore, based on the Cr concentrations and isotopic composition of the Earth's crust and chondrites, we varied the amount of chondritic impactor and tested different Cr concentrations for the target rock, and modeled their mixing (Fig. 4). The goal of this calculation was to verify if the Cr signature of our samples could be reproduced. In the case of PR9, its $^{54}$Cr deficit is compatible with an admixture of 2-3% of an ordinary chondrite component ($\varepsilon^{54}$Cr =-0.40; Cr=3600 ppm) diluted in continental crust material with $\varepsilon^{54}Cr = 0$ and $Cr = 90$ ppm (Rudnick and Gao., 2003). In consequence, we suggest that an ordinary chondrite component is recorded in Sudbury ejecta.

Finally, the Cr isotope composition of C9 also needs to be discussed, especially because it is different from PR and C18 samples. The coupled excess of $^{53}Cr$ and $^{54}Cr$ ($\varepsilon^{53}Cr=0.19\pm0.04$ and $\varepsilon^{54}Cr=0.26\pm0.07$) is in general characteristic of carbonaceous chondrites. However, mixing calculations attempts reveal that, although the $\varepsilon^{54}Cr$ composition of C9 can easily be reproduced by mixing only a few percent (~1%) of any carbonaceous chondrite type with terrestrial component it is not possible to match the $\varepsilon^{53}Cr$ composition using the same amount of impactor. Significantly higher proportions (~10%) of meteoritic material are required to account for the $\varepsilon^{53}Cr$ values of the sample (Fig. S2). Its composition does not represent a simple binary mixture between any of the known meteorite types and terrestrial material.

To summarize, among the four samples measured for Cr isotopes in Sudbury ejecta deposits, at least two present Cr isotopic anomalies (PR9 and C9). However, both have opposite signs. PR10 shows no Cr isotope difference compared to Earth but has the highest PGE abundances, which could also be indicative of an enstatite chondrite supplementary component. The record of multiple meteoritic components in different samples could simply suggest the implication of multiple impactors from different nature. However, it would be in contradiction with all the arguments previously detailed in the section 2 supporting a single impact origin for the Sudbury ejecta deposits from Lake superior area (Cannon et al. 2010), as well as it still does not explain the peculiar composition of C9.

Another way to reconcile our results with the idea of a single impact event is to consider a mixed meteorite impactor for the Sudbury event. The existence of heterogeneities inside impactors has previously been described in the case of the Almahata Sitta object, which consists of multiple fragments of different chondrites (H, EH, EL) and ureilites (Bischoff et al., 2010). Heterogeneous impactors appear to be very rare given the lack of observable evidence for heterogeneities within a single object even from the asteroid belt. However, current observations may just not be able to detect fine-scale heterogeneities. Therefore, in the absence of clear geological evidence for multiple impacts, the hypothesis of a heterogeneous impactor currently represents the only option that satisfies all the observations. A multi-component impactor with local melting and mixing of its components prior or during the impact with the terrestrial target is also more consistent with the complex signature of C9.

### 4.2. Grænsesø and Zaonega spherule layers

Both pairs of samples from the Grænsesø (GL-1 and GL-5) and the Zaonega (FD27.36 and FD26.51) spherule layers exhibit evident Cr isotopic anomalies that now clearly confirm the presence of extra-terrestrial material in both spherule layers. Their positive $\varepsilon^{54}$Cr signature is similar to carbonaceous chondrite composition (Fig. 1). The samples also define a correlation in a $\varepsilon^{53}$Cr vs $\varepsilon^{54}$Cr plot (Fig. 2) that points towards the field of carbonaceous chondrites (more specifically CI or CR chondrites). This trend falls out of the field of Lake Superior ejecta deposits (Fig. 2), and therefore unambiguously invalidates a possible relationship with the Sudbury event, leaving the Vredefort impact event or a not yet discovered (or totally eroded) crater the only candidates eligible to the formation of these ejecta deposits. This correlation could also represent a mixing line between the Earth and carbonaceous chondrites suggesting that these two spherule layers are possibly co-genetic. GL-5 and FD27.36 fall on different mixing curves in Fig. 4, however, both samples can also result from the impact of a similar type of chondritic object (CI or CR) provided the target rock shows some variability in Cr concentration.

The Cr composition of FD27.36 can be explained by mixing ~4% of a CI component ($\varepsilon^{54}$Cr=1.7, Cr=2650 ppm) with 96% of a terrestrial target having the following composition: $\varepsilon^{54}$Cr =0, Cr=40 ppm. For comparison, the Cr concentration measured in the non-spherulitic parts of the drill core that sampled FD samples varies from 40 to 60 ppm (Huber et al., 2014a). CI-type material also matches the composition of GL-5 if 1.5% of a uniform impactor is diluted in a more enriched crustal target rock (Cr ~160 ppm). Although a CR-like impactor ($\varepsilon^{54}$Cr=1.3, Cr=3415 ppm) is also possible in the case of GL-5, it is most unlikely for FD27.36 because it would require a terrestrial component with a too low Cr concentration ($\leq$5 ppm). Variations in the Cr concentrations of the target rock are to be expected for large continental impacts, as they would affect large volume of rock of contrasted lithologies compared to smaller and/or marine impacts. This is for example the case for the Cr concentrations measured in the target rocks from the Vredefort structure, which range from ~10 to 750 ppm (Koeberl et al., 2002). On the contrary, K-T boundary ejecta related to the Chixculub impact event show a single and strong correlation between Cr isotopes and concentrations (Trinquier et al., 2006) that can be explained by the relative homogeneity of target rock composed of marine sediments.

In summary, the samples from the Grænsesø and Zaonega spherule layers all contain positive Cr isotopic anomalies, which represent solid evidence for meteoritic material into these two layers. Their Cr isotopic compositions are clearly distinct from those of Pine River and Coleraine ejecta, which makes the Sudbury impact event impossible to be at their origin, and prefers the Vredefort or unknown impact events. The samples are also correlated with each other in $\varepsilon^{53}$Cr - $\varepsilon^{54}$Cr space (Fig. 2), which could suggest a single major impact event as the source of the Grænsesø and Zaonega spherule layers. This trend unambiguously points towards the field of carbonaceous chondrites, most likely CI or CR types. However, because the Cr isotopic compositions of CI and CR are very close, the possibility that both layers are related to two different impactors from each type (or even same type) cannot be totally excluded either.

## 5. CONCLUSION

We analyzed the Cr isotope composition of samples from the only three discovered Paleoproterozoic layers that are likely to be of impact origin. Lake Superior ejecta were previously attributed to the Sudbury impact event (Cannon et al., 2010; Huber et al., 2014b), whereas the origins of the Zaonega and the Grænsesø spherule layers remain unknown. Unlike the Grænsesø layer, the Zaonega layer was previously confirmed as related to a chondritic object fall, but the nature of this impactor was not resolved. Our new results confirm the presence of extra-terrestrial material in all layers, and therefore their impact origin. Based on their positive Cr isotopic anomalies, we propose that both spherule layers contain carbonaceous chondrite component (CI and possibly CR). The fact that GL and FD samples form a trend in $\varepsilon^{53}$Cr - $\varepsilon^{54}$Cr isotopic plot could suggest a common origin, however, two different impactors cannot definitely be ruled out. We also show that the Cr isotopic signature of the Grænsesø and Zaonega spherule layers is unambiguously different from Lake Superior ejecta deposits that are related to the Sudbury event. Finally, the multiple isotopic signatures observed in Lake Superior rocks are complex and remain difficult to interpret. They are incompatible with a classical model of a uniform and unique impactor. Therefore, we suggest that the Sudbury impactor was most probably a mixed meteorite that was composed of multiple chondritic components. Although it would be unusual because rare, this is currently the only option that satisfies new and previous results. The involvement of cometary impactors (Pope et al., 2004; Petrus et al., 2015) can be neither excluded nor confirmed by our Cr isotope results because the Cr isotopic composition of comets is unknown, although, the Stardust missions found that cometary dust contains carbonaceous chondrites particles (e.g., Nakashima et al., 2012).

## ACKNOWLEDGMENTS

We thank B. Glass and F. Smith for the GL samples, M. Huber for the PR and C samples, and the ICDP FAR-DEEP project for the FD samples. Editor C. Sotin and the two anonymous reviewers are thanked for the time spent on the manuscript and their interesting comments. We thank J. L. Birck for his insightful advice and his help during the analytical work. S. Goderis, B. Kamber, and B. Simonson are also thanked for their helpful comments on this paper. FM thanks the ERC under the European Community's H2020 framework program/ERC grant agreement # 637503 (Pristine), the ANR for a chaire d'Excellence Sorbonne Paris Cité (IDEX13C445), the UnivEarthS Labex program (ANR-10-LABX-0023 and ANR-11-IDEX-0005-02), the IPGP multidisciplinary program PARI, and the Region île-de-France SESAME for Grant no. 12015908.

*Accepted 7[th] December 2016 EPSL*

and Deutsch, A., eds., Large meteorite impacts III. Geological Society of America, pp. 1-24. Special Paper 384. doi: 10.1130/0-8137-2384-1.1.

**FIGURE CAPTIONS**

Table 1 Chromium isotope compositions ($\varepsilon^{54}$Cr and $\varepsilon^{53}$Cr) and siderophile element concentrations of bulk ejecta samples from North America (PR & C), Russia (FD), Greenland (GL) and of BHVO-2 terrestrial sample (*data from Huber et al., 2014b; **data from Huber et al., 2014a; ***data from Chadwick et al. 2001). The uncertainties of individual Cr isotopic analyses are two standard errors (2se). The letter *n* indicates the number of replicates. The notation $\varepsilon^{53}$Cr and $\varepsilon^{54}$Cr represents the relative deviation in parts per 10,000 of $^{53}$Cr/$^{52}$Cr and $^{54}$Cr/$^{52}$Cr ratios from the terrestrial standard.

Figure 1 – $\varepsilon^{54}$Cr values measured in Sudbury ejecta (Pine River and Coleraine drill cores), and Zaonega and Grænsesø spherule deposits (see legend of Table 1 and text for details) compared to eucrites, enstatite, ordinary, and carbonaceous chondrites (Meteorite data compilation from Göpel et al., 2015).

Figure 2 – $\varepsilon^{53}$Cr versus $\varepsilon^{54}$Cr plot. Comparison of Pine River, Coleraine, Zaonega and Grænsesø ejecta samples (rectangles) compared to the average compositions (adapted from Foriel et al., 2013) of chondrites (circles), eucrite (dotted circle) and the Earth (dashed rectangles).

Figure 3 – CI-normalized Platinum Group Element abundance patterns for samples from a) Zaonega (Huber et al., 2014a) and Grænsesø (Chadwick et al., 2001) spherule layers, and for the Vredefort granophyre (Huber et al., 2014a), b) Pine River and Coleraine ejecta deposits (Huber et al., 2014b), and average composition of the Sudbury Igneous Complex (Mungall et al., 2004).

Figure 4 – $\varepsilon^{54}$Cr versus Cr concentration diagram. The curved lines represent mixing lines between different proportions (%) of carbonaceous chondrites (CI, grey lines and full circles; CR, grey dashed lines and crosses) and ordinary chondrites (OC, black line and squares), and terrestrial target rocks (Cr= 5, 40, 90, and 160 ppm). Large triangles represent the compositions of Zaonega and Grænsesø ejecta (FD27.36 and GL-5 respectively), and large circles, the composition of Sudbury ejecta (C9, C18, PR9 and PR10). Error bars are smaller than the size of the symbols.

**Table 1**

| Samples | n | $\varepsilon^{54}Cr$ | 2 s.e | $\varepsilon^{53}Cr$ | 2 s.e | Cr (ppm) | Co (ppm) | Ni (ppm) | Ir (ppb) | Ru (ppb) | Pt (ppb) | Rh (ppb) | Pd (ppb) | Au (ppb) |
|---|---|---|---|---|---|---|---|---|---|---|---|---|---|---|
| *Sudbury ejecta layer* | | | | | | | | | | | | | | |
| PR9 | 5 | -0.20 | 0.09 | -0.04 | 0.05 | 188* | 4.3* | 54.2* | 0.96* | 2.10* | 4.49* | 0.76* | 7.9* | 0.46* |
| PR10 | 6 | -0.01 | 0.08 | 0.03 | 0.04 | 223 | 14.1 | 62.8 | 1.16* | 2.68* | 13.4* | 1.11* | 9.36* | 3.18* |
| C9 | 10 | 0.26 | 0.07 | 0.19 | 0.04 | 222 | 20.9 | 172* | 0.86* | 2.69* | 2.18* | 0.94* | 3.55* | 1.47* |
| C18 | 2 | 0.05 | 0.09 | 0.02 | 0.03 | 30.7* | 15.7* | 32.5* | 0.21* | 0.5* | 2.45* | 0.22* | 4.06* | 0.72* |
| *Zaonega spherule layer* | | | | | | | | | | | | | | |
| FD27.36 | 2 | 1.26 | 0.07 | 0.18 | 0.04 | 152 | 156 | 1339 | 0.14** | 0.44** | 2.33** | 1.25** | 7.08** | 7.76** |
| FD26.51 | 6 | 1.06 | 0.11 | 0.14 | 0.06 | n.a | n.a | n.a | n.a | n.a | n.a | n.a | n.a | n.a |
| *Grænsesø spherule layer* | | | | | | | | | | | | | | |
| GL1 | 6 | 0.54 | 0.11 | 0.14 | 0.05 | n.a | n.a | n.a | n.a | n.a | n.a | n.a | n.a | n.a |
| GL5 | 5 | 0.35 | 0.09 | 0.07 | 0.04 | 200 | 12.2 | 131 | n.a | n.a | n.a | n.a | n.a | n.a |
| GGU | | n.a | n.a | n.a | n.a | 92*** | 24*** | 149*** | 0.02*** | 0.93*** | 0.12*** | 0.73*** | 2.07*** | 0.25*** |
| *Terrestrial rock standard* | | | | | | | | | | | | | | |
| BHVO-2 | 5 | 0.03 | 0.10 | 0.02 | 0.06 | | | | | | | | | |

**Figure 1**

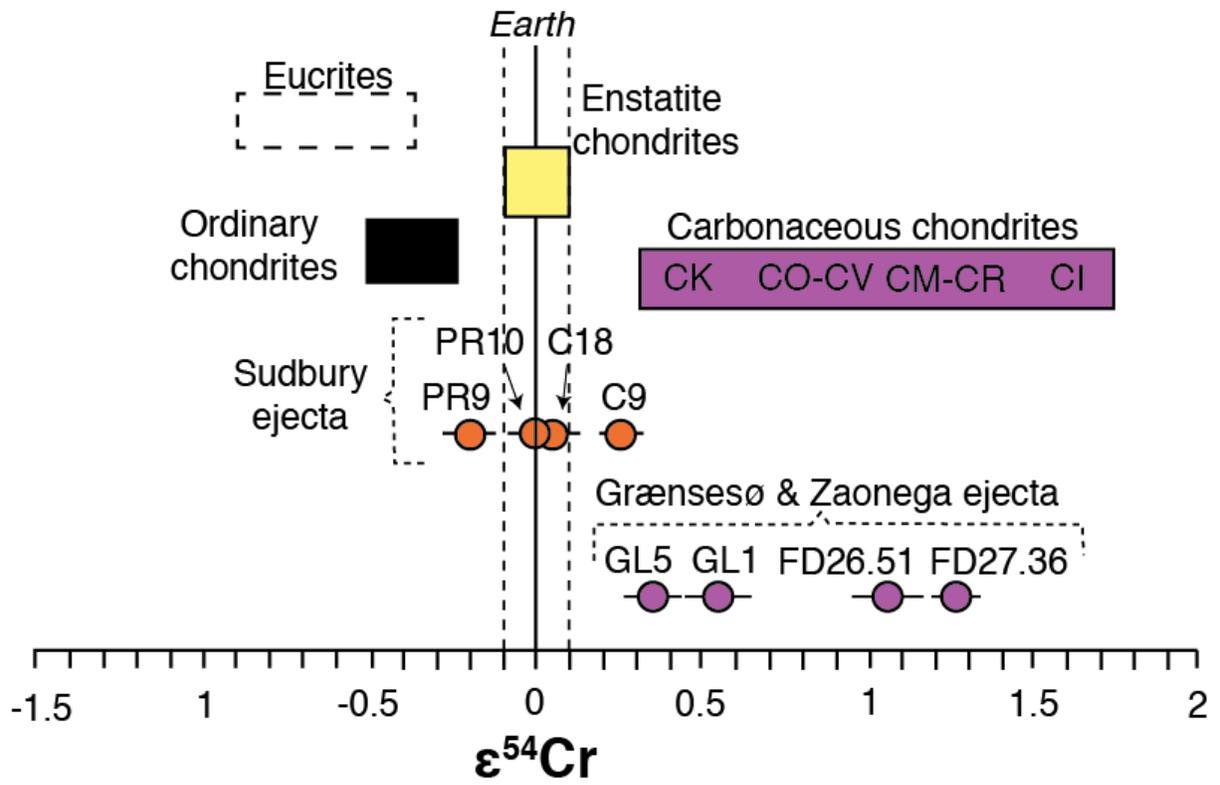

**Figure 2**

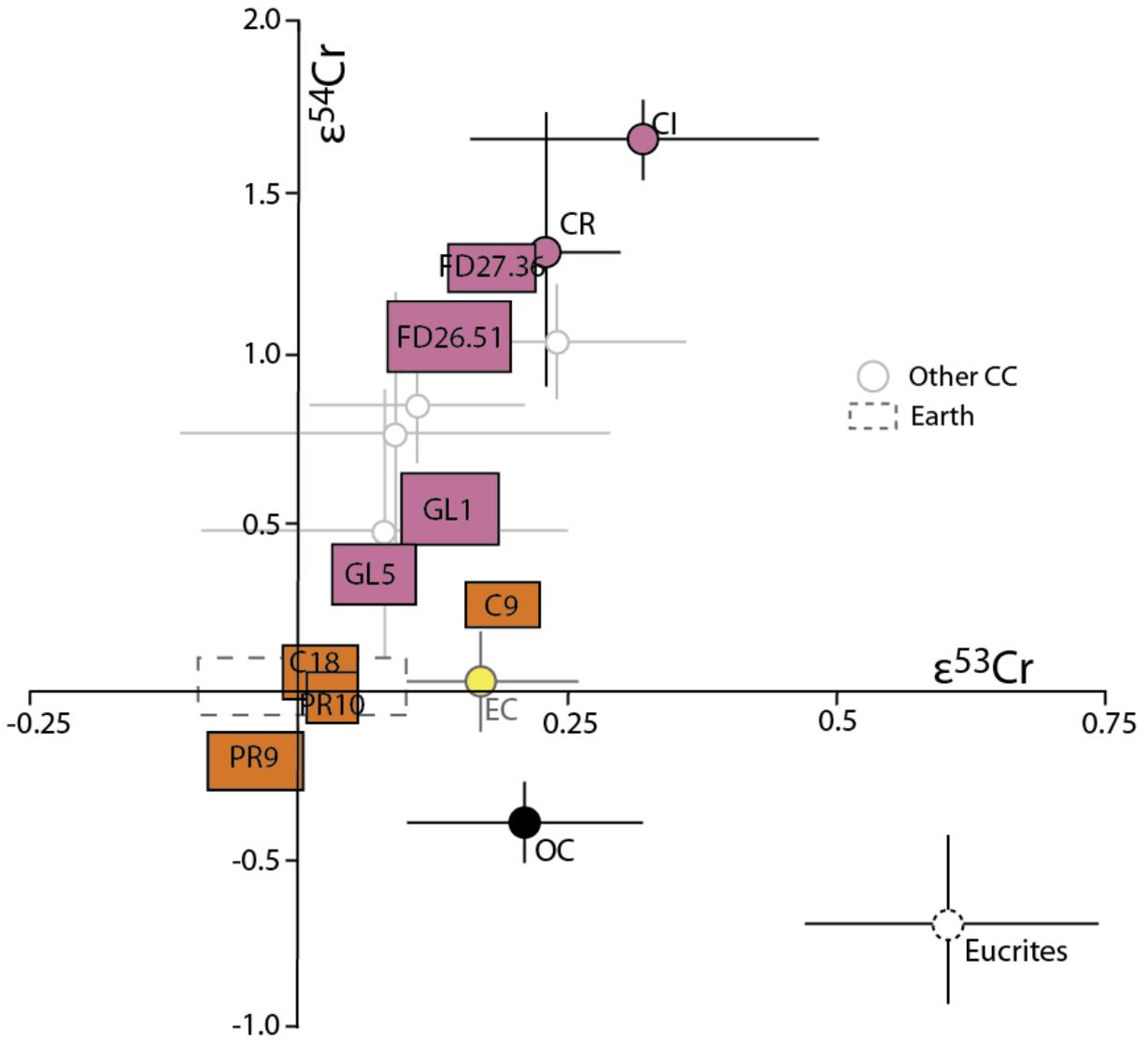

**Figure 3**

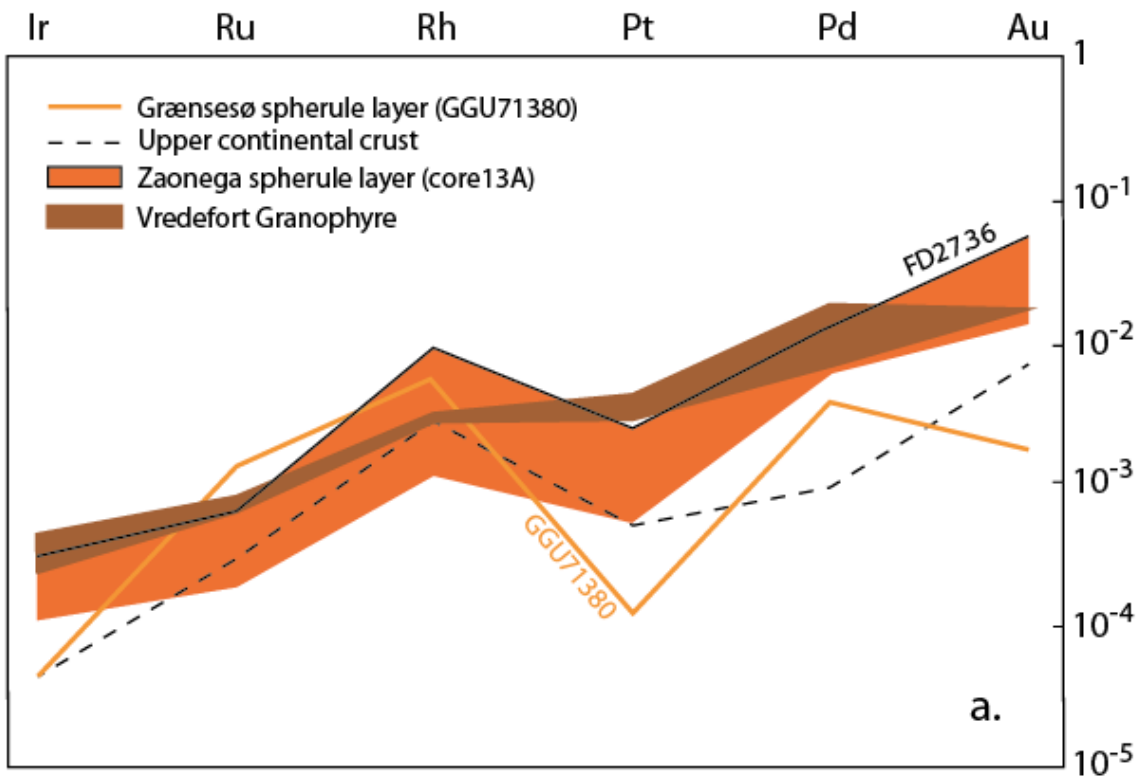

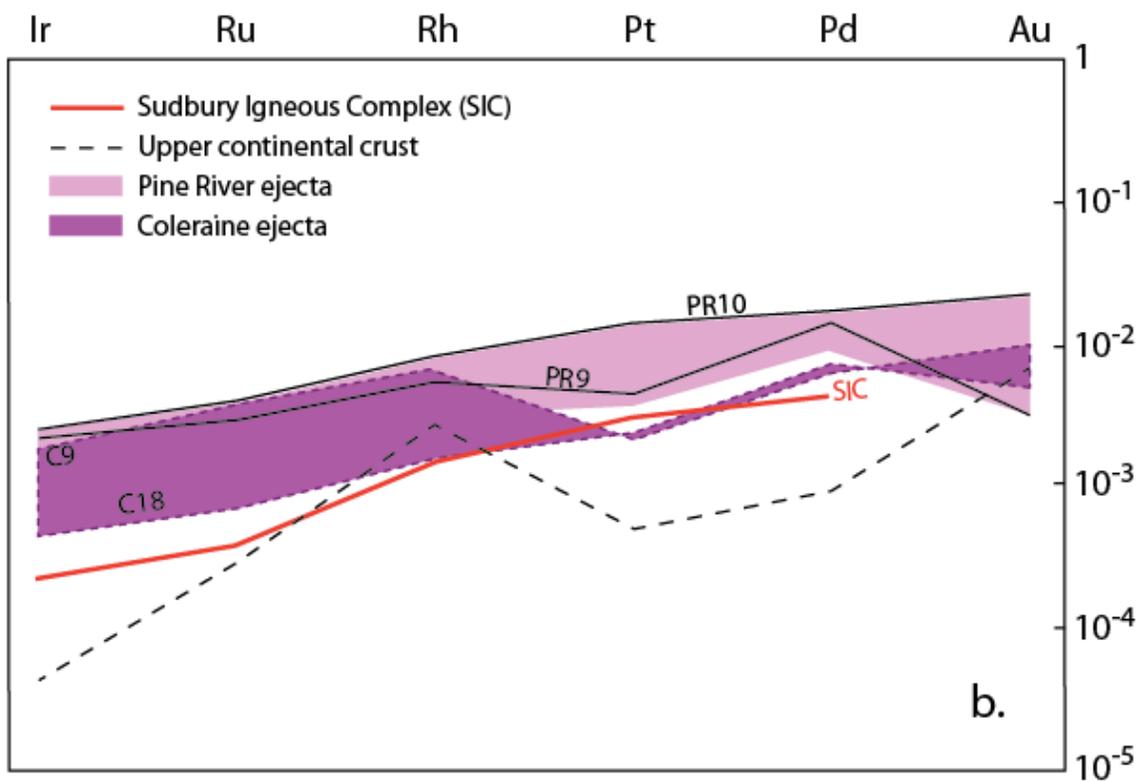

**Figure 4**

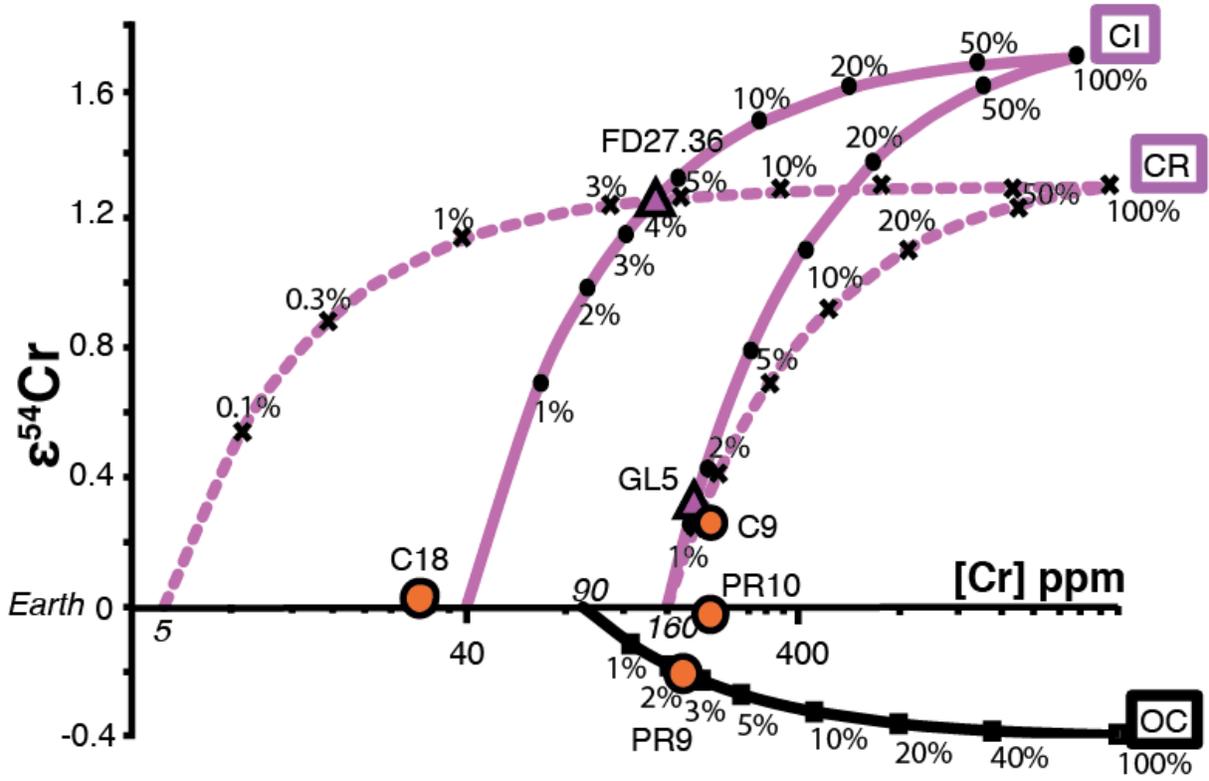

SUPPLEMENTARY MATERIAL

For *Chromium isotope evidence in ejecta deposits for the nature of Paleoproterozoic impactors* by Mougel et al.

**SUPPLEMENTARY FIGURE AND TABLE CAPTIONS**

Figure S1 – Idealized stratigraphic columns of the Mesabi Range, Gunflint Range, and Sudbury structure showing the generally accepted correlations between these 3 areas, including the Pine River and Coleraine ejecta, and Sudbury crater impact rocks. Adapted from Cannon et al. (2010).

Figure S2 – $\varepsilon^{54}Cr$ and $\varepsilon^{53}Cr$ versus Cr concentration diagram. The curved lines represent mixing lines between carbonaceous chondrites (CI) and terrestrial material. Open and full symbols are associated with $\varepsilon^{54}Cr$ and $\varepsilon^{53}Cr$ curves, respectively. This diagram shows that it is impossible to reproduce $\varepsilon^{54}Cr$ and $\varepsilon^{53}Cr$ compositions of C9 with the same amount of impactor material. While only 1% of impactor component is required to match $\varepsilon^{54}Cr$ value, almost 10% is needed to fit $\varepsilon^{53}Cr$ value. On the contrary, mixing lines for GL-5 show that the same amount of CI impactor (1-2%) can easily explain both the $\varepsilon^{54}Cr$ and $\varepsilon^{53}Cr$ compositions.

Table S1 – Cr elution procedure.

Table S2 – Multi-line cup configuration for Cr isotopic measurements.

Figure S1

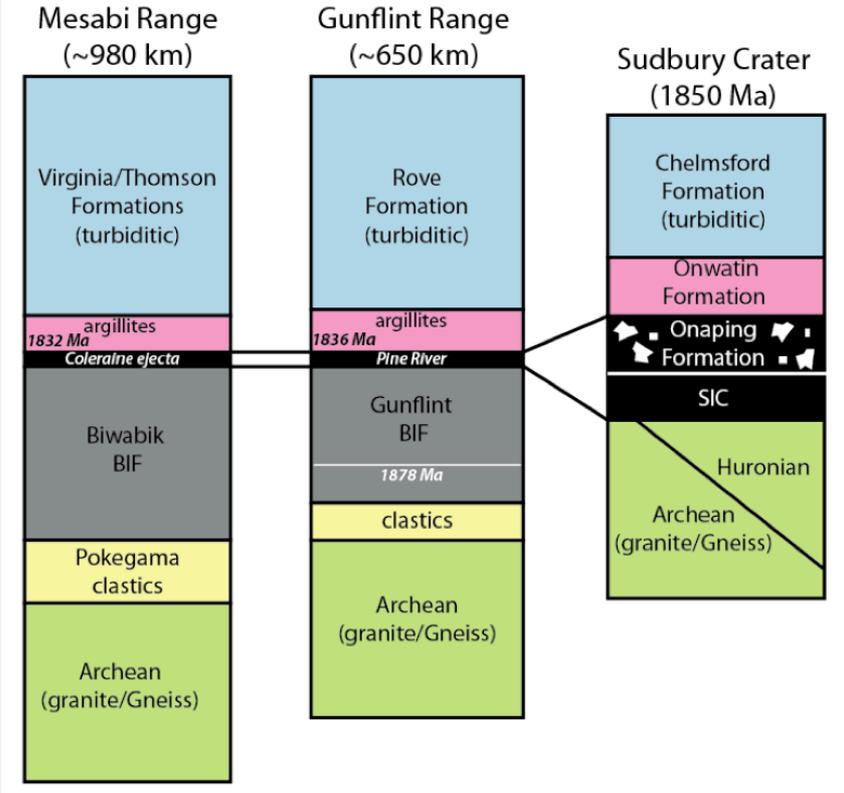

Figure S2

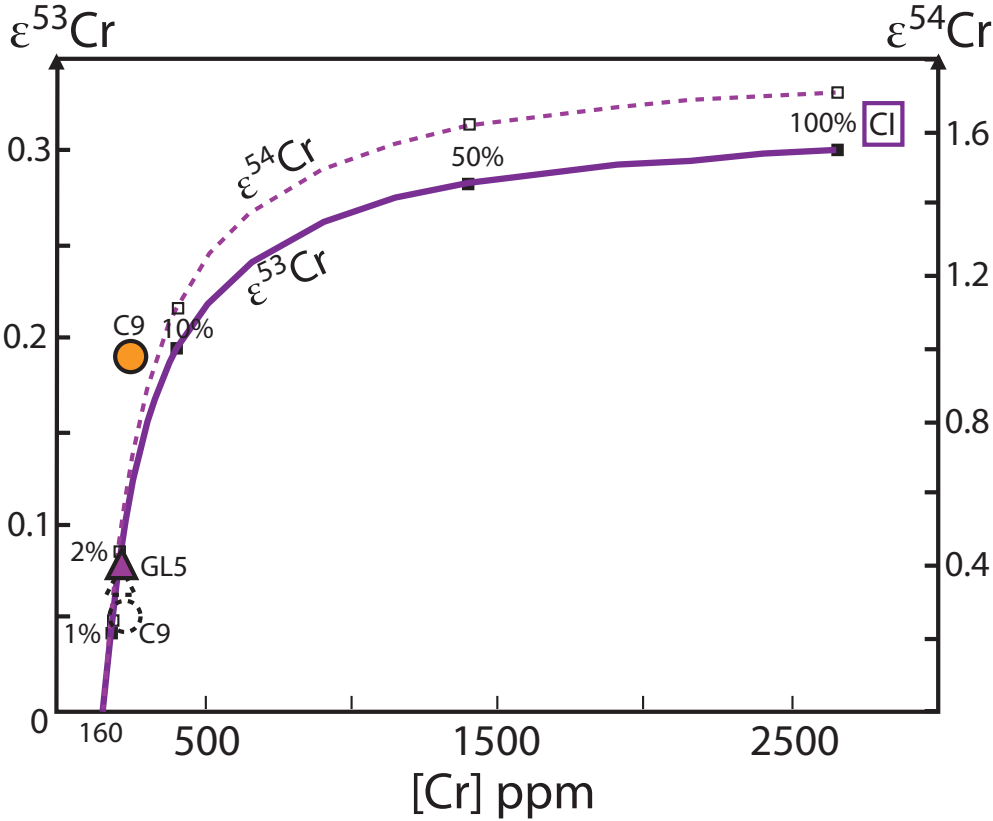

Table S1

| Elution 1 | 1mL 50WX8 200-400 mesh | |
|---|---|---|
| wash | 6N HCL - H20 - 6N HCl | |
| conditioning | H2O | 1 mL |
| **Introduction + Collection 1** | ~1N HCl | 1.1 mL |
| Rinse (collection1) | 1N HCl | 1 mL |
| Rinse (collection1) | 1N HCl | 1.5 mL |
| Rinse (collection1) | 1N HCl | 2.5 mL |
| **Collection 2*** | 2N HCl | 4 mL |
| * Elution1 repeated on "collection 2" fraction for yield better than ≥99% | | |
| **Elution 2** | **0.3mL 50WX8 200-400 mesh** | |
| wash | 6N HCL - H20 - 6N HCl | |
| conditioning | H2O | 1 mL |
| **Introduction** | ~ 0.8N HNO3 | 2.1 mL |
| Rinse | 0.8N HF | 1 mL |
| Rinse | 0.8N HF | 1.5 mL |
| Rinse | 1N HCl | 0.5 mL |
| Rinse | 1N HCl | 1 mL |
| Rinse | 1N HCl | 1.5 mL |
| Rinse | 1N HCl | 4.5 mL |
| **Collection** | 2N HCl | 3 mL |

Table S2

| Line | L4 | L3 | L2 | L1 | C | H1 | H2 | H3 | H4 |
|------|-----|-----|-----|-----|-----|-----|-----|-----|-----|
| 1 | 49 | $^{50}$Cr | 51 | $^{52}$Cr | $^{53}$Cr | $^{54}$Cr | 55 | 56 | 57 |
| 2 | 48 | 49 | $^{50}$Cr | 51 | $^{52}$Cr | $^{53}$Cr | $^{54}$Cr | 55 | 56 |
| 3 | $^{50}$Cr | 51 | $^{52}$Cr | $^{53}$Cr | $^{54}$Cr | 55 | 56 | 57 | 58 |